\pdfoutput=1

\documentclass[11pt,twoside,a4paper,cmspaper,final,collab]{cms-tdr}
\def\svnVersion{486182P}\def\cmsCernNoTag{CERN-EP-2018-183}\def\cmsCernDate{\today}\def\cmsMessage{Published in Physical Review Letters as \href{http://dx.doi.org/10.1103/PhysRevLett.122.011803}{\doi{10.1103/PhysRevLett.122.011803.}}}

\begin{document}\cmsNoteHeader{EXO-16-049}

\hyphenation{had-ron-i-za-tion}
\hyphenation{cal-or-i-me-ter}
\hyphenation{de-vices}
\RCS$HeadURL: svn+ssh://svn.cern.ch/reps/tdr2/papers/EXO-16-049/trunk/EXO-16-049.tex $
\RCS$Id: EXO-16-049.tex 486015 2019-01-11 15:49:27Z ksung $

\newlength\cmsFigWidth
\ifthenelse{\boolean{cms@external}}{\setlength\cmsFigWidth{0.336\textwidth}}{\setlength\cmsFigWidth{0.45\textwidth}}
\ifthenelse{\boolean{cms@external}}{\providecommand{\cmsUpperLeft}{upper\xspace}}{\providecommand{\cmsUpperLeft}{upper left\xspace}}
\ifthenelse{\boolean{cms@external}}{\providecommand{\cmsLowerRight}{lower\xspace}}{\providecommand{\cmsLowerRight}{lower\xspace}}
\ifthenelse{\boolean{cms@external}}{\providecommand{\cmsUpperRight}{middle\xspace}}{\providecommand{\cmsUpperRight}{upper right\xspace}}
\ifthenelse{\boolean{cms@external}}{\providecommand{\cmsLeft}{upper\xspace}}{\providecommand{\cmsLeft}{left\xspace}}
\ifthenelse{\boolean{cms@external}}{\providecommand{\cmsRight}{lower\xspace}}{\providecommand{\cmsRight}{right\xspace}}

\newcommand{\Znunu}{\ensuremath{\cPZ(\nu\overline{\nu})\!+\!\text{jets}}\xspace}
\newcommand{\Wlnu}{\ensuremath{\PW(\ell\nu)\!+\!\text{jets}}\xspace}
\newcommand{\Wjets}{\ensuremath{\PW\!+\!\text{jets}}\xspace}
\newcommand{\Zjets}{\ensuremath{\cPZ\!+\!\text{jets}}\xspace}
\newcommand{\Vjets}{\ensuremath{\cmsSymbolFace{V}\!+\!\text{jets}}\xspace}
\newcommand{\gq}{\ensuremath{\Pg_{\cPq}}\xspace}
\newcommand{\gqq}{\ensuremath{\Pg_{{\cPq\cPq}}\xspace}}
\newcommand{\ttDM}{\ensuremath{\ttbar\!+\!\chi\overline{\chi}}\xspace}
\newcommand{\mwt}{\ensuremath{\mTii^{\PW}}\xspace}
\newcommand{\mtll}{\ensuremath{\mTii^{\ell\ell}}\xspace}

\newcommand{\ptvecjet}{\ensuremath{\ptvec^{\kern1pt\mathrm{j}}}\xspace}
\newcommand{\ptvecjetonetwo}{\ensuremath{\ptvec^{\kern1pt\mathrm{j_{1,2}}}}\xspace}
\newcommand{\mindphi}{\ensuremath{\min\Delta\phi(\ptvecjet,\ptvecmiss)}}
\newcommand{\mindphionetwo}{\ensuremath{\min\Delta\phi(\ptvecjetonetwo,\ptvecmiss)}}

\cmsNoteHeader{EXO-16-049}

\title{Search for dark matter particles produced in association with a top quark pair at \texorpdfstring{$\sqrt{s} = 13\TeV$}{sqrt(s) = 13 TeV}}

\date{\today}

\abstract{
A search is performed for dark matter particles produced in association with a top quark pair in proton-proton collisions at $\sqrt{s} = 13\TeV$. The data correspond to an integrated luminosity of 35.9\fbinv recorded by the CMS detector at the LHC. No significant excess over the standard model expectation is observed. The results are interpreted using simplified models of dark matter production via spin-0 mediators that couple to dark matter particles and to standard model quarks, providing constraints on the coupling strength between the mediator and the quarks. These are the most stringent collider limits to date for scalar mediators, and the most stringent for pseudoscalar mediators at low masses.
}

\hypersetup{%
pdfauthor={CMS Collaboration},%
pdftitle={Search for dark matter produced in association with a top quark pair at sqrt(s) = 13 TeV},%
pdfsubject={CMS},%
pdfkeywords={CMS, physics, dark matter}}

\maketitle

Astrophysical observations strongly motivate the existence of dark matter~\cite{Bertone:2004pz,Feng:2010gw,Porter:2011nv,1475-7516-2018-03-026}, which may originate from physics beyond the standard model.  In a large class of models, dark matter consists of stable, weakly interacting massive particles ($\chi$)~\cite{1475-7516-2018-03-026}, which may be pair produced at the CERN LHC via mediators that couple both to dark matter particles and to standard model quarks.  The dark matter particles would escape detection, thereby creating a transverse momentum imbalance ($\ptvecmiss$) in the event.  Searches at collider experiments can offer insights on the nature of the mediator and provide constraints on dark matter masses of O($10\GeV$) and below, a region that is difficult to explore both in direct and indirect searches for dark matter.  A favored class of models proposes a spin-0 mediator with standard model Higgs-like Yukawa coupling to quarks, which therefore couples preferentially to the top quark~\cite{Haisch:2012kf,Lin:2013sca,Buckley:2014fba,Haisch:2015ioa,Arina:2016cqj}.  Consequently, in this class of models dark matter production in association with a top quark pair ($\ttbar$) can offer better search sensitivity compared to other modes such as production in association with a jet~\cite{Aad:2015zva,Khachatryan:2016mdm,Aaboud:2016tnv,Sirunyan2017,EXO16048}.  At the LHC, the $\ttDM$ process is probed through the signature of $\ttbar$ accompanied by $\ptvecmiss$~\cite{Khachatryan:2015nua,Aad:2014vea}.

The top quark almost always decays to a {\PW} boson and a {\cPqb} quark. The {\PW} boson can decay leptonically (to a charged lepton and a neutrino) or hadronically (to a quark pair).  The signal regions (SRs) of the search cover three $\ttbar$ decay modes: the all-hadronic, lepton+jets ($\ell+$jets where $\ell=\Pe,\mu$), and dileptonic ($\Pe\Pe$, $\Pe\mu$, $\mu\mu$) final states where neither, either, or both of the {\PW} bosons decay to leptons, respectively.  This Letter presents a search for $\ttDM$ in {\Pp\Pp} collisions at $\sqrt{s}=13\TeV$ with data recorded by the CMS experiment in 2016, corresponding to an integrated luminosity of 35.9\fbinv.  The analysis strategy is similar to Ref.~\cite{EXO16005}, but includes additional SRs for the dileptonic mode.

The central feature of the CMS detector is a superconducting solenoid providing a magnetic field of 3.8\unit{T}.  Within the solenoid volume are the silicon pixel and strip trackers, a lead tungstate crystal electromagnetic calorimeter, and a brass and scintillator hadron calorimeter.  A steel and quartz-fiber Cherenkov forward hadron calorimeter extends the pseudorapidity ($\eta$) coverage.  The muon system consists of gas-ionization detectors embedded in the steel flux-return yoke outside the solenoid.  A two-tiered trigger system~\cite{cmstrigger} selects events at a rate of about 1\unit{kHz} for storage.  A detailed description of the CMS detector is provided in Ref.~\cite{CMS}.

The event reconstruction is based on the CMS particle-flow algorithm~\cite{CMS-PRF-14-001}, which reconstructs and identifies individual particles using an optimized combination of the detector information.  The $\ptvecmiss$ vector is computed as the negative vector sum of the transverse momenta ($\ptvec$) of all the particles in an event.  Jets are formed from particles using the anti-\kt algorithm~\cite{Cacciari:2008gp,Cacciari:2011ma} with a distance parameter of 0.4.  Corrections are applied to calibrate the jet momentum~\cite{2011JInst...611002C} and to remove energy from additional collisions in the same or adjacent bunch crossings (pileup)~\cite{Cacciari:2008gn}.  Jets in the analysis are required to have $\pt>30\GeV$ and $\abs{\eta}<2.4$, and to satisfy identification criteria~\cite{CMS-PAS-JME-16-003} that minimize spurious detector effects.  A combined secondary vertex {\cPqb} tagging algorithm~\cite{btag} is used to identify jets originating from {\cPqb} quarks (\cPqb-tagged jets).  A multivariate discriminant, the ``resolved top tagger'' (RTT)~\cite{EXO16005}, based on jet properties and kinematic information, is used to identify top quarks that decay into three jets.  Electrons and muons are selected using ``tight'' and ``loose'' requirements where the former applies more stringent identification criteria than the latter~\cite{Khachatryan:2015hwa,Khachatryan:2015hwa}.  The tight leptons are used in the selection of specific final states, while loose leptons are used to veto events with extra leptons.  The primary {\Pp\Pp} interaction vertex is taken to be the reconstructed vertex with the highest summed $\pt^2$ of its associated physics objects.  Here, the physics objects are the jets, clustered with the tracks assigned to the vertex as inputs, plus the associated $\ptvecmiss$.

The $\ttDM$ signal results in high-$\pt$ jets including \cPqb-tagged jets, leptons, and significant $\ptvecmiss$.  The background is dominated by $\ttbar$ and $\Vjets$ ($\text{V}=\PW,\cPZ/\gamma^{*}$) production.  The $\ttbar$ and single top quark backgrounds are simulated at next-to-leading order (NLO) accuracy in quantum chromodynamics (QCD) using \POWHEG v2 and \POWHEG v1~\cite{Nason:2004rx,Frixione:2007vw,Alioli:2010xd,Oleari:2010nx}, respectively.  Samples of $\Vjets$ and QCD multijet events are simulated at leading order (LO) in QCD using \MGvATNLO~v2.2.2~\cite{Alwall:2014hca} (\MADGRAPH), with up to four additional partons in the matrix element (ME) calculations.  The $\Vjets$ samples are corrected with boson $\pt$-dependent electroweak corrections~\cite{Denner:2009gj, Denner:2011vu, Denner:2012ts, Kuhn:2005gv, Kallweit:2014xda, Kallweit:2015dum} and QCD NLO/LO $K$ factors computed using \MADGRAPH.  Samples of $\ttbar\!+\!\text{V}$ and diboson processes (\PW\PW, \PW\cPZ, and {\cPZ}{\cPZ}) are generated at NLO using either \MADGRAPH or \POWHEG v2.  The initial-state partons are modeled with the NNPDF 3.0~\cite{Ball:2014uwa} parton distribution function (PDF) sets at LO or NLO in QCD to match the ME calculation.  Generated events are interfaced with \PYTHIA 8.205~\cite{Sjostrand:2007gs} for parton showering using the \textsc{CUETP8M1} tune~\cite{Khachatryan:2015pea}, except for the $\ttbar$ simulation which uses the \textsc{CUETP8M2} tune customized by CMS with an updated strong coupling $\alpS$ for initial-state radiation.  The simulation of the CMS detector is performed with \GEANTfour~\cite{Agostinelli:2002hh}.  Corrections derived from data are applied to account for any mismodeling of selection efficiencies in simulation.

The signal is simulated using simplified models of dark matter production~\cite{Abercrombie:2015wmb}.  The dominant mechanism is $s$-channel production of the mediator via gluon fusion, with the mediator then decaying to a pair of dark matter particles.  The dark matter particles are assumed to be Dirac fermions, and the mediators are spin-0 particles with scalar ($\phi$) or pseudoscalar ($\Pa$) interactions.  The couplings between the mediator and standard model quarks are $\gqq=\gq y_{\cPq}$, where $y_{\cPq}=\sqrt{2}m_{\cPq}/v$ are the standard model Yukawa couplings, $m_{\cPq}$ is the quark mass, and $v=246\GeV$ is the Higgs boson field vacuum expectation value.  The $\gq$ parameter is assumed to be unity for all quarks.  The direct coupling strength of the mediators to dark matter is denoted by $\Pg_{\chi}$.  The model does not take into account possible mixing between $\phi$ and the standard model Higgs boson~\cite{ALBERT201749}.  The $\ttDM$ signal is generated at LO using \MADGRAPH with up to one additional parton, and the mediator is forced to decay to a pair of dark matter fermions.  The mediator width is computed according to partial-width formulas in Ref.~\cite{PhysRevD.91.055009} and assuming no additional interactions beyond those described here.  The relative width of the scalar (pseudoscalar) mediator varies between 4\% and 6\%\,(4\% and 8\%) for masses in the range of 10--500\GeV.  The signal is normalized to the cross section computed at NLO in QCD.

Data are collected by triggering on events containing large $\ptmiss$ (the magnitude of $\ptvecmiss$) or high-\pt leptons.  The triggers for the all-hadronic final state are based on the amount of $\ptmiss$ and \mht measured with the trigger-level reconstruction.  The \mht variable is defined as the magnitude of the vector sum of $\ptvec$ over trigger-level jets with $\pt>20\GeV$ and $\abs{\eta}<5.0$ that pass identification requirements.  During the period of data collection, the $\ptmiss$ and \mht trigger thresholds were increased as the instantaneous luminosity increased, in steps from 90 to 120\GeV.  Events in the $\ell+$jets final state are obtained using single-lepton triggers that require an electron (muon) with $\pt>27$ (24)\GeV.  Events in the dilepton final state are obtained using single-lepton and dilepton ($\Pe\Pe$, $\Pe\mu$, $\mu\mu$) triggers.  The trigger thresholds on the higher- and lower-\pt electrons (muons) are 23\,(17)\GeV and 12\,(8)\GeV, respectively, and apply to all pairings of lepton object flavors.

Using additional selection requirements, two all-hadronic, one $\ell+$jets, and four dilepton SRs are defined.  Several control regions (CRs) enriched in standard model processes are used to improve the simulation-based background estimates for the all-hadronic and $\ell+$jets SRs.  There are no event overlaps among the regions.  Together, the SRs and CRs associated with the individual $\ttDM$ final states are referred to as ``channels''.  All three $\ttDM$ channels are used in a simultaneous maximum-likelihood fit of $\ptmiss$ distributions to extract a potential dark matter signal.  In the fit, the CRs constrain the contributions of $\ttbar$, $\Wjets$, and $\Zjets$ processes within each channel via freely floating normalization parameters for each $\ptmiss$ bin.

The all-hadronic SRs require $\ptmiss>200\GeV$, and four or more jets, of which at least one must be {\cPqb} tagged.  Any event with a loose lepton of $\pt>10\GeV$ is vetoed.  The dominant background consists of $\ttbar$ decays to $\ell+$jets, referred to as $\ttbar(1\ell)$, where the lepton is not identified as loose and therefore not vetoed, and the neutrino is the source of $\ptvecmiss$.  The RTT is employed to define a category of events with two tagged hadronic top quark decays (2RTT), which suppresses the $\ttbar(1\ell)$ background, and a category with less than two top quark tags (0,1RTT) and at least two \cPqb-tagged jets.  The RTT variable is essentially independent of $\ptmiss$; therefore, any bias on the $\ptmiss$ shape from requiring top tags is negligible.  Spurious $\ptmiss$ can arise in multijet events as a result of jet energy mismeasurement.  In such cases, the reconstructed $\ptvecmiss$ tends to align with a jet.  The multijet background is suppressed by requiring the smallest azimuthal angle between the $\ptvecmiss$ and each jet in the event, $\Delta_{\mathrm{j}}\equiv\mindphi$, to be greater than $0.4$ ($1.0$) radians in the 2RTT (0,1RTT) category.  The $\Delta_{\mathrm{j}}$ requirement also reduces the $\ttbar(1\ell)$ background, for which $\ptvecmiss$ can align with a bottom jet.

The CRs targeting the $\ttbar(1\ell)$ background, one for each category, are defined by selecting events with exactly one tight lepton with $\pt>30\GeV$, and by requiring the transverse mass, \mT, given in terms of $\ptvecmiss$ and the lepton momentum ($\ptvec^{\ell}$) by the following expression:
\begin{linenomath*}
\begin{equation}
\mT=\sqrt{\smash[b]{2\pt^{\ell}\ptmiss[1-\cos\Delta\phi(\ptvec^{\ell},\ptvecmiss)]}},
\end{equation}
\end{linenomath*}
to be less than 160\GeV, in order to avoid overlaps with the SR of the $\ell+$jets channel.  For ideal measurements, the \mT quantity is bounded above by the {\PW} boson mass for $\ttbar(1\ell)$ and for $\Wlnu$ where the {\PW} boson is produced on shell.

There are also significant background contributions from $\Znunu$, and from $\Wlnu$ where the lepton is not identified.  The CRs enriched in both $\Wjets$ and $\Zjets$ are formed by modifying the SR selections to require no \cPqb-tagged jets.  Additionally, dedicated $\Wjets$ CRs are defined by requiring a tight lepton with $\pt>30\GeV$ and $\mT<160\GeV$.  A CR enriched in $\Zjets$ is defined by selecting two tight, oppositely charged, and same-flavor leptons, with the dilepton invariant mass between 60 and 120\GeV.  This CR is not subdivided into categories based on the number of top quark tags because the event yield with two tags is negligible. The $\ptmiss$ calculation does not consider the two leptons in order to emulate the $\Znunu$ process.

Events in the $\ell+$jets SR are selected by requiring $\ptmiss>160\GeV$, exactly one tight lepton with $\pt>30\GeV$, and three or more jets, of which at least one is {\cPqb} tagged.  Events must not contain additional loose leptons with $\pt>10\GeV$.  A selection of $\mT>160\GeV$ is imposed to reduce the $\ttbar(1\ell)$ and $\Wjets$ backgrounds.  Following these selections, the remaining background events primarily consist of dileptonic $\ttbar$ decays, referred to as $\ttbar(2\ell)$ events, where one of the leptons is not identified.  This background is suppressed by requiring that the $\mwt$ quantity~\cite{MT2W} be larger than 200\GeV, and for the two highest \pt jets in the event that $\Delta_{\mathrm{j}_{1,2}}\equiv\mindphionetwo>1.2$.  The $\mwt$ variable corresponds to the minimum mass of a particle consistent with being pair-produced and decaying to a bottom quark and a {\PW} boson, where both {\PW} bosons decay leptonically but one of the two leptons is not detected.  The key characteristic of $\mwt$ is that for ideal measurements the distribution for $\ttbar(2\ell)$ events is bounded above by the top quark mass.

The CR targeting the $\ttbar(2\ell)$ background is defined by requiring an additional tight lepton of $\pt>30\GeV$ with respect to the SR selection and removing the selections on \mT, \mwt, and $\Delta_{\mathrm{j}_{1,2}}$.  To reduce the signal contamination and avoid overlap with the dileptonic SRs, the $\mtll$ variable~\cite{Lester:1999tx,Burns:2008va,Cheng:2008hk} is required to be less than 110\GeV.  The $\mtll$ variable is essentially the minimum \mT of a pair-produced particle that decays to a lepton and a neutrino.  The $\mtll$ distribution for $\ttbar(2\ell)$ events is bounded above by the {\PW} boson mass for ideal measurements.  A \Wjets~CR for the $\ell+$jets channel is defined by requiring no \cPqb-tagged jets and removing the selections on $\mwt$ and $\Delta_{\mathrm{j}_{1,2}}$.

Events in the dilepton SRs are selected by requiring exactly two oppositely charged tight leptons with higher (lower) $\pt>25$ (15)\GeV, two or more jets with at least one \cPqb-tagged jet, and $\ptmiss>50\GeV$.  The dilepton mass is required to be greater than 20\GeV, and for the dielectron and dimuon events, to be at least 15\GeV away from the {\cPZ} boson pole mass ($m_\cPZ$)~\cite{pdg}.  Separate categories are considered for events with same- and different-flavor lepton pairs, and for events with $\mtll$ greater or less than 110\GeV.  Additionally, events with $\mtll<110\GeV$ must not pass the selection for the $\ttbar(2\ell)$ CR of the $\ell+$jets search channel.  The SRs with large $\mtll$ have significantly higher signal purity.

The estimates for the backgrounds from Drell--Yan production and from jets misidentified as leptons are performed using dedicated sideband regions not included in the fit.  A $\ptmiss$-dependent correction to the Drell--Yan simulation in the same-flavor SRs is obtained by comparing data and simulation yields within 15\GeV of $m_\cPZ$.  The misidentified leptons background is estimated using data events with pairs consisting of one tight lepton plus a ``nontight'' lepton-like object passing a less stringent selection.  The number of such combinations is scaled by misidentification rates, which are measured in a jet-enriched sample.

The dark matter signal, which would be observed as an excess of events compared to the predicted background at high $\ptmiss$, is extracted via a simultaneous maximum-likelihood fit to the binned $\ptmiss$ distributions of the signal and backgrounds, based on simulation with the exception of the misidentified leptons background, in the SRs and associated CRs. The fit is performed using the \textsc{RooStats} statistical package~\cite{RooStats}.  The template shapes and normalization are allowed to vary in the fit, constrained by the priors of the systematic uncertainties, parametrized as nuisance parameters.

The common sources of uncertainty are correlated across SRs and CRs and across channels.  The sources of normalization uncertainty include the integrated luminosity (2.5\%)~\cite{CMS-PAS-LUM-17-001}, {\cPqb} tagging efficiency (1\%--5\%)~\cite{btag}, lepton efficiency (1.5\%--2.2\%)~\cite{Khachatryan:2015hwa,Khachatryan:2015hwa}, and pileup simulation (0.8\%--1.9\%), where the range of values indicates variations across different physics processes.  The common sources of shape uncertainty include $\ptmiss$ trigger efficiency, jet energy scale and resolution~\cite{CMS-PAS-JME-16-003}, PDF~\cite{Ball:2014uwa}, uncertainty on the $K$ factors for $\Vjets$, uncertainty from missing higher order QCD corrections for each simulated physics process, and the uncertainty in the modeling of top quark $\pt$ in $\ttbar$ simulation~\cite{PhysRevD.95.092001}.  The jet energy scale uncertainties have the largest impact on the $\ell+$jets and dilepton channels, while in the all-hadronic channel the top quark $\pt$ modeling and $\ptmiss$ trigger uncertainties are more important.

Within the all-hadronic and $\ell+$jets search channels, additional nuisance parameters scale the yields of the $\ttbar$, $\Wjets$, and $\Zjets$ backgrounds independently in each $\ptmiss$ bin across the SRs and CRs of a given channel.  For example, in each bin of $\ptmiss$ a single parameter is associated with the contribution of the $\Wjets$ process in the all-hadronic SRs and CRs, while another set of parameters distinct from those of the all-hadronic channel, is associated with the $\Wjets$ background in the $\ell+$jets SRs and CRs.  These nuisance parameters allow the data in the CRs to constrain the estimates of the dominant background processes in the corresponding SRs.  Signal yields in all the SRs and CRs are scaled simultaneously by the signal strength parameter ($\mu$), defined as the ratio of the measured signal cross section to the theoretical cross section, $\mu=\sigma/\sigma_{\text{th}}$.

\begin{figure}[h!!tb]
\centering
  \includegraphics[width=\cmsFigWidth]{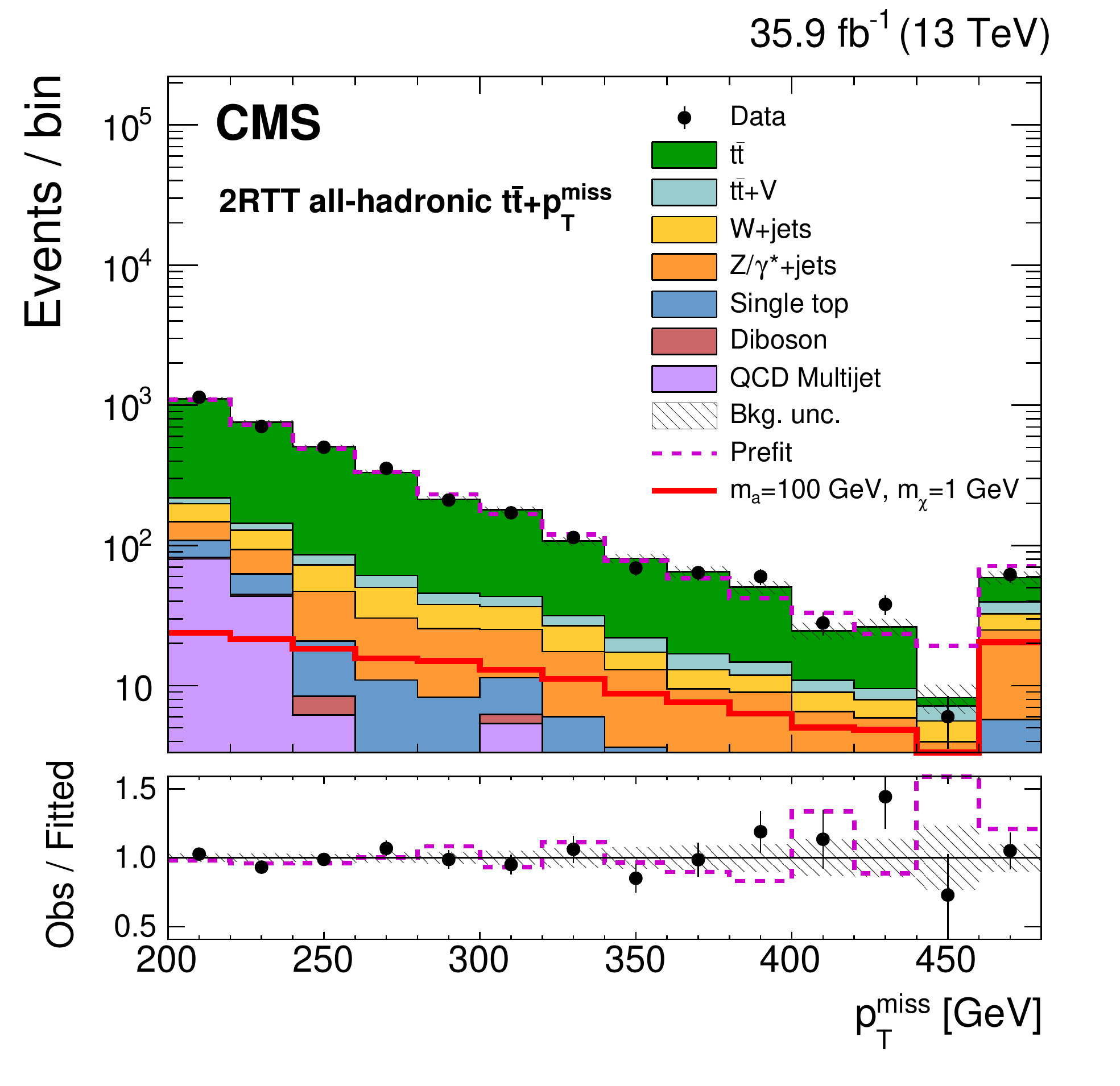}
  \includegraphics[width=\cmsFigWidth]{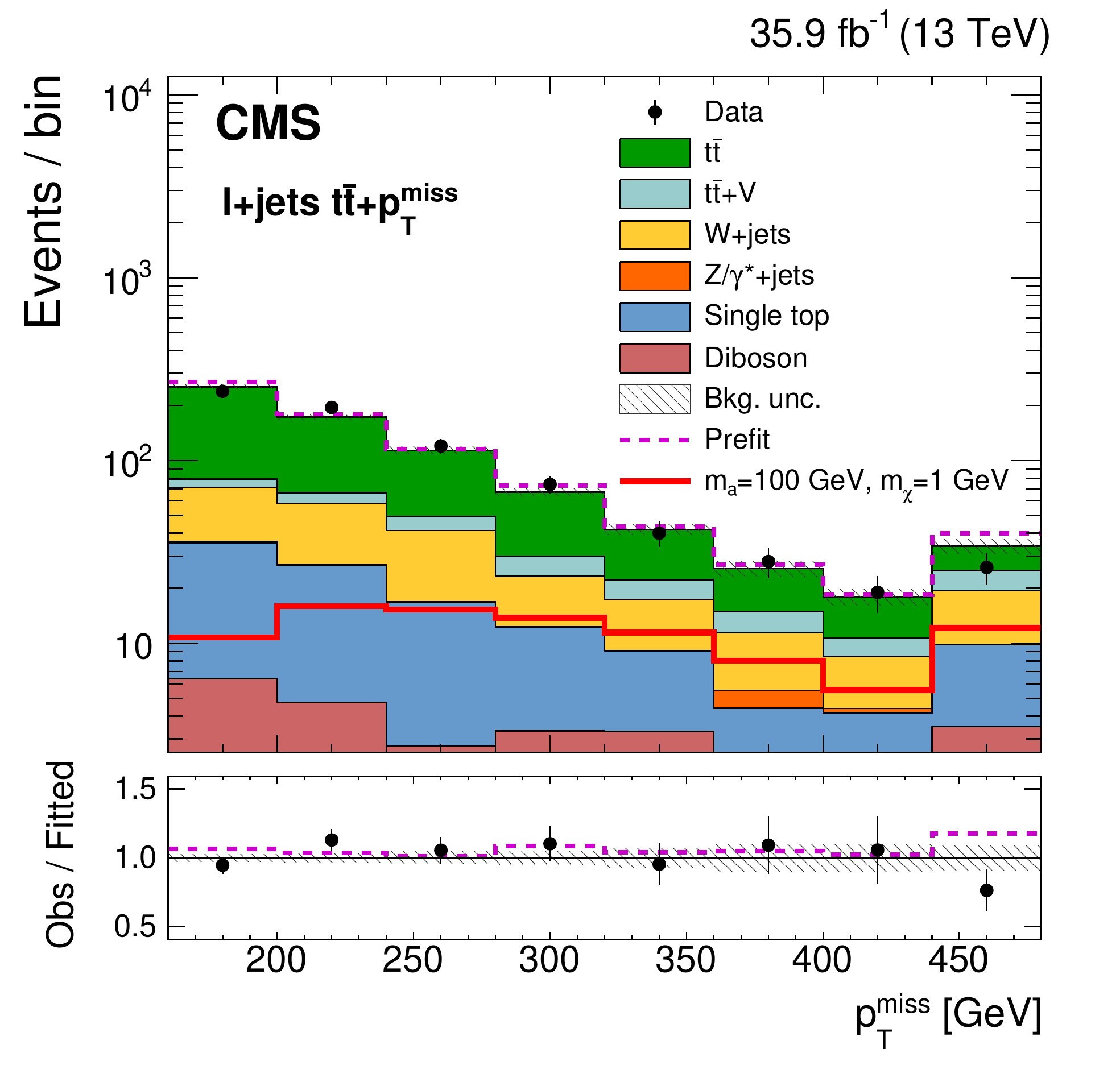}
  \includegraphics[width=\cmsFigWidth]{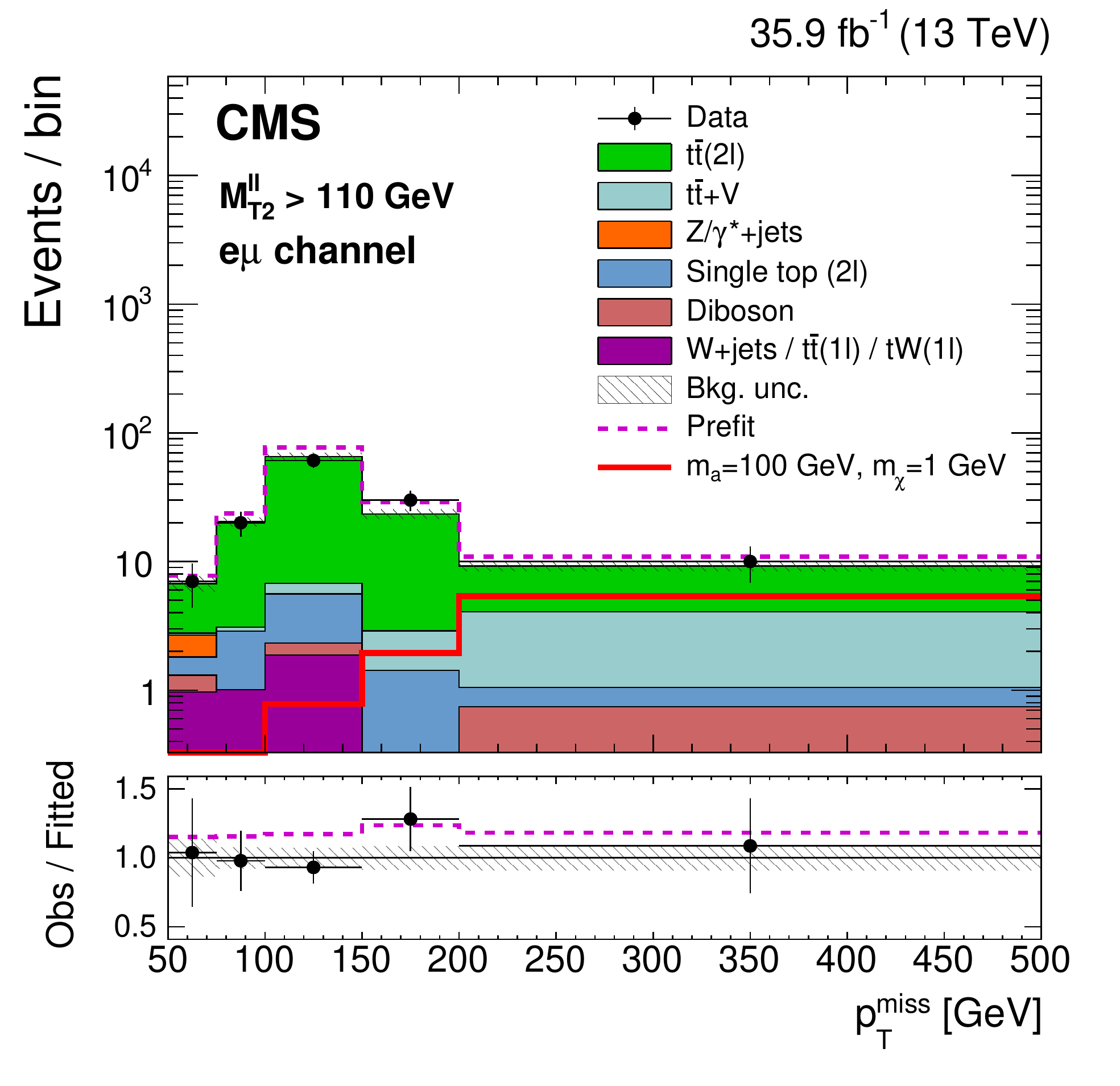}
  \caption{Selected $\ptmiss$ distributions in SRs: 2RTT SR for the all-hadronic (\cmsUpperLeft), the $\ell+$jets (\cmsUpperRight), and the different-flavor, $\mtll>110\GeV$ SR in the dileptonic channel (\cmsLowerRight).  The solid red line shows the expectation for a signal with $m_\Pa = 100\GeV$ and $m_{\chi} = 1\GeV$.  The last bin contains the overflow events.  The lower panel shows the ratio of the observed to the fitted distribution (points), and the ratio of the background expectation before the fit to the fitted distribution (dashed magenta line).  The vertical bars indicate the statistical uncertainty on the data.  The horizontal bars on the rightmost plot indicate the bin width.  The uncertainty bands in both panels include the statistical and systematic uncertainties on the total background.}
  \label{fig:postfit_ptmiss}
\end{figure}

The fit is performed across all search channels and no significant excess is observed.  Figure~\ref{fig:postfit_ptmiss} shows the $\ptmiss$ distributions for three of the seven SRs, obtained after the background-only fit assuming the absence of any signal.  Upper limits are set on the $\ttDM$ production cross section using a modified frequentist approach (\CLs) with a test statistic based on the profile likelihood in the asymptotic approximation~\cite{JUNK1999435,cls,Cowan2011}.  For each signal hypothesis, 95\% confidence level (\CL) upper limits on $\mu$ are determined.  The all-hadronic channel provides the best sensitivity.  The dileptonic channel is competitive with the all-hadronic channel for scalar mediator masses less than about 50\GeV, where the signal has a soft $\ptmiss$ spectrum, but is typically the least sensitive channel in other regions of the parameter space.

\begin{figure}[h!tb]
\centering
  \includegraphics[width=0.45\textwidth]{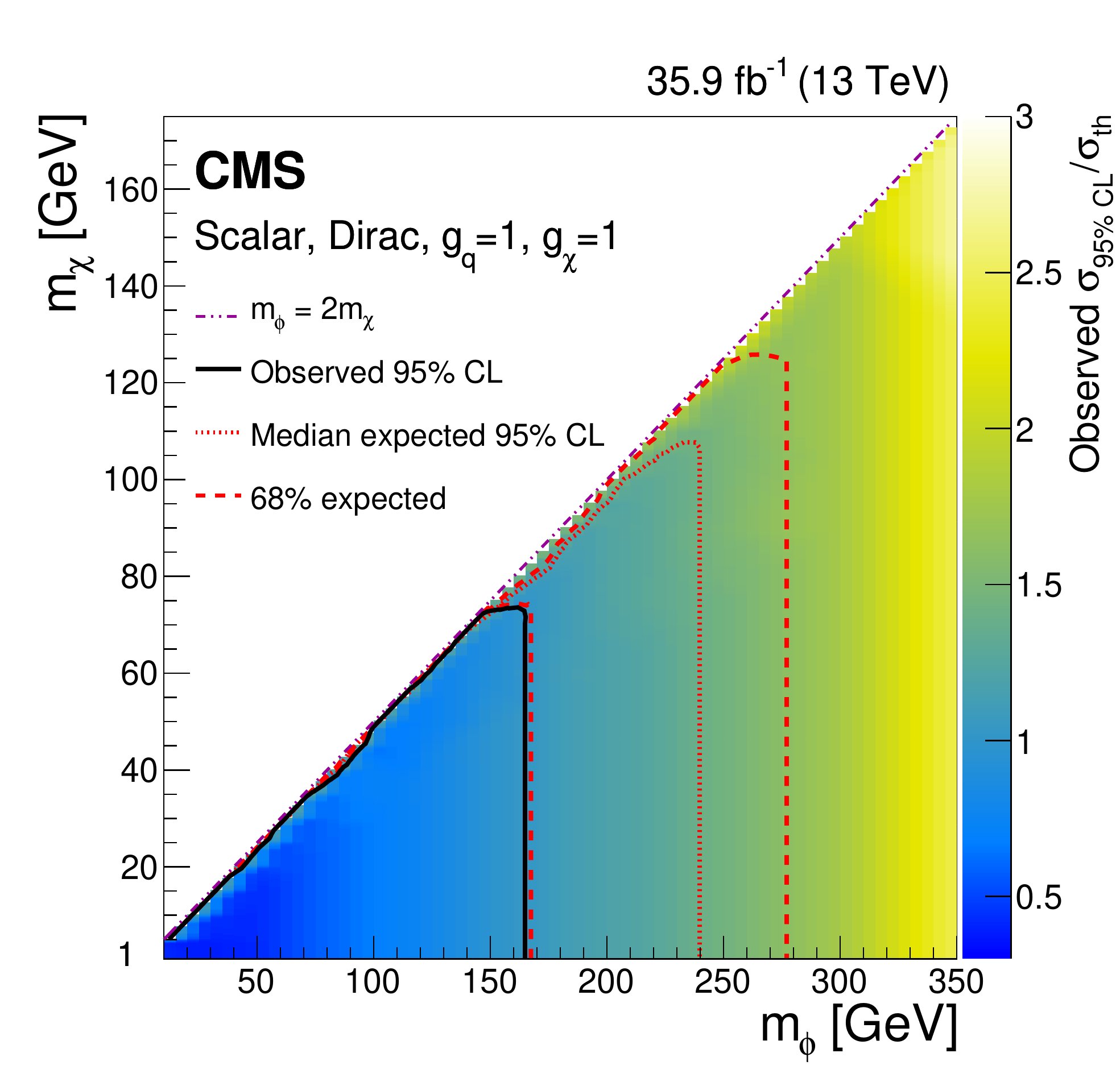}
  \includegraphics[width=0.45\textwidth]{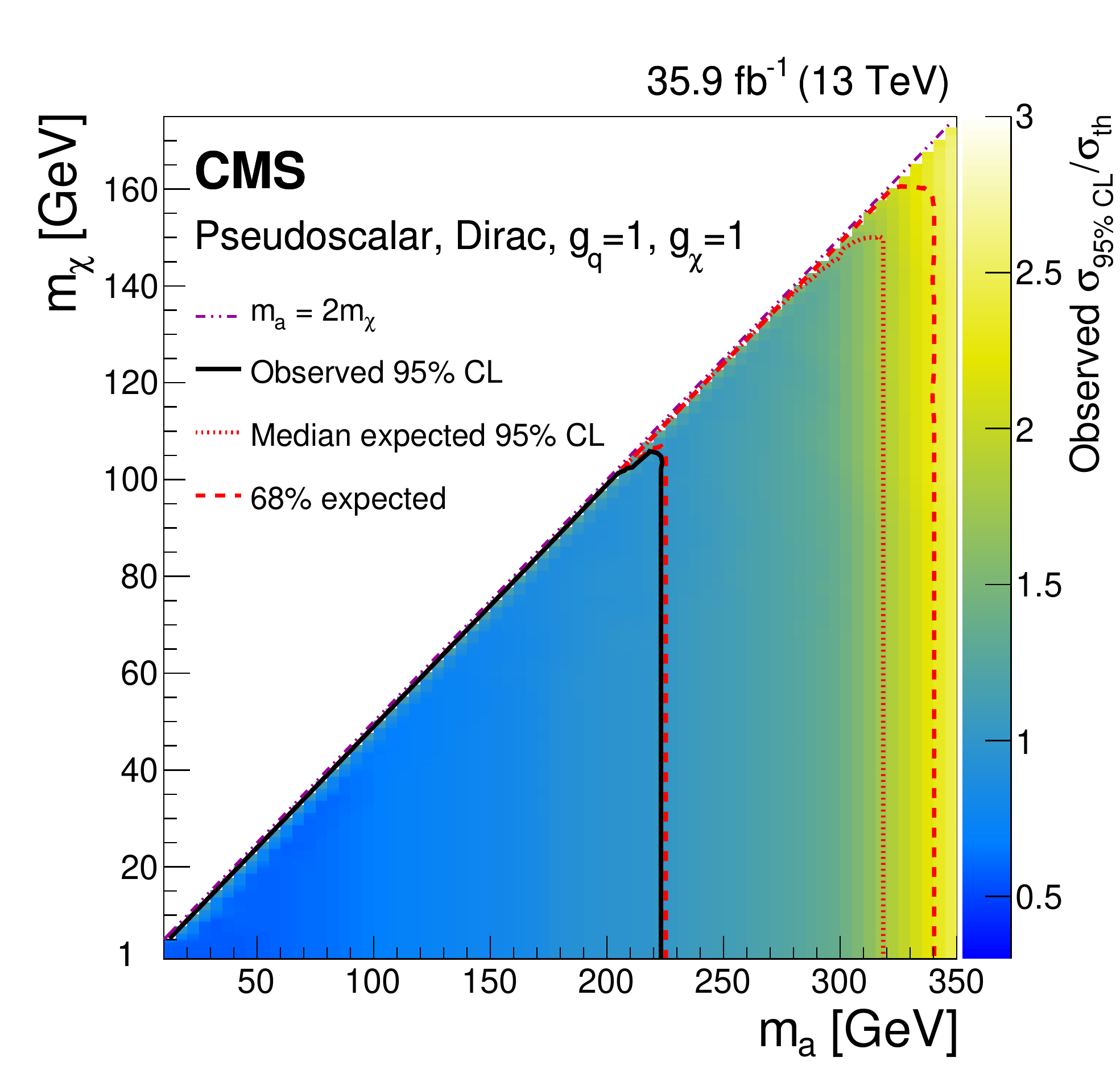}
  \caption{The exclusion limits at 95\% \CL on the signal strength $\mu=\sigma/\sigma_{\text{th}}$ computed as a function of the mediator and dark matter mass, assuming a scalar (\cmsLeft) and pseudoscalar (\cmsRight) mediator. The mediator couplings are assumed to be $\gq=\Pg_{\chi}=1$.  The dashed magenta lines represent the 68\% probability interval around the expected limit.  The observed limit contour is almost coincident with the boundary of the 68\% probability interval.
}
  \label{fig:limits_mdm_mmed_combined}
\end{figure}

The limits are shown as a function of $m_{\Pa/\phi}$ and $m_\chi$ in Fig.~\ref{fig:limits_mdm_mmed_combined}.  The contours enclose the region where the upper limit on $\mu$ is less than 1.  Because of the narrow width of the mediator, the signal cross section drops rapidly across the $m_{\Pa/\phi} = 2m_\chi$ line, marking the boundary between the on-shell to the off-shell region. Therefore, the exclusion contour runs close to the $m_{\Pa/\phi} = 2m_\chi$ line but does not cross it.  The observed (expected) upper limits on $\mu$ exclude scalar and pseudoscalar masses of 160\,(240) and 220\,(320)\GeV, respectively, at 95\% \CL.  The observed exclusion is weaker than the expected because of tension in the fit between CRs and SRs of the all-hadronic channel, although the difference is not significant as the observed result lies only just outside the 68\% probability interval.  This arises because the a priori estimation of the background exceeds the number of events observed in the CRs, while the estimate is in better agreement with data in the SRs.  Consequently, the signal+background fit, in contrast to the background-only fit, reduces this tension between CRs and SRs by accommodating for some signal, which contributes primarily to the SRs.

\begin{figure}[h!tb]
\centering
  \includegraphics[width=0.45\textwidth]{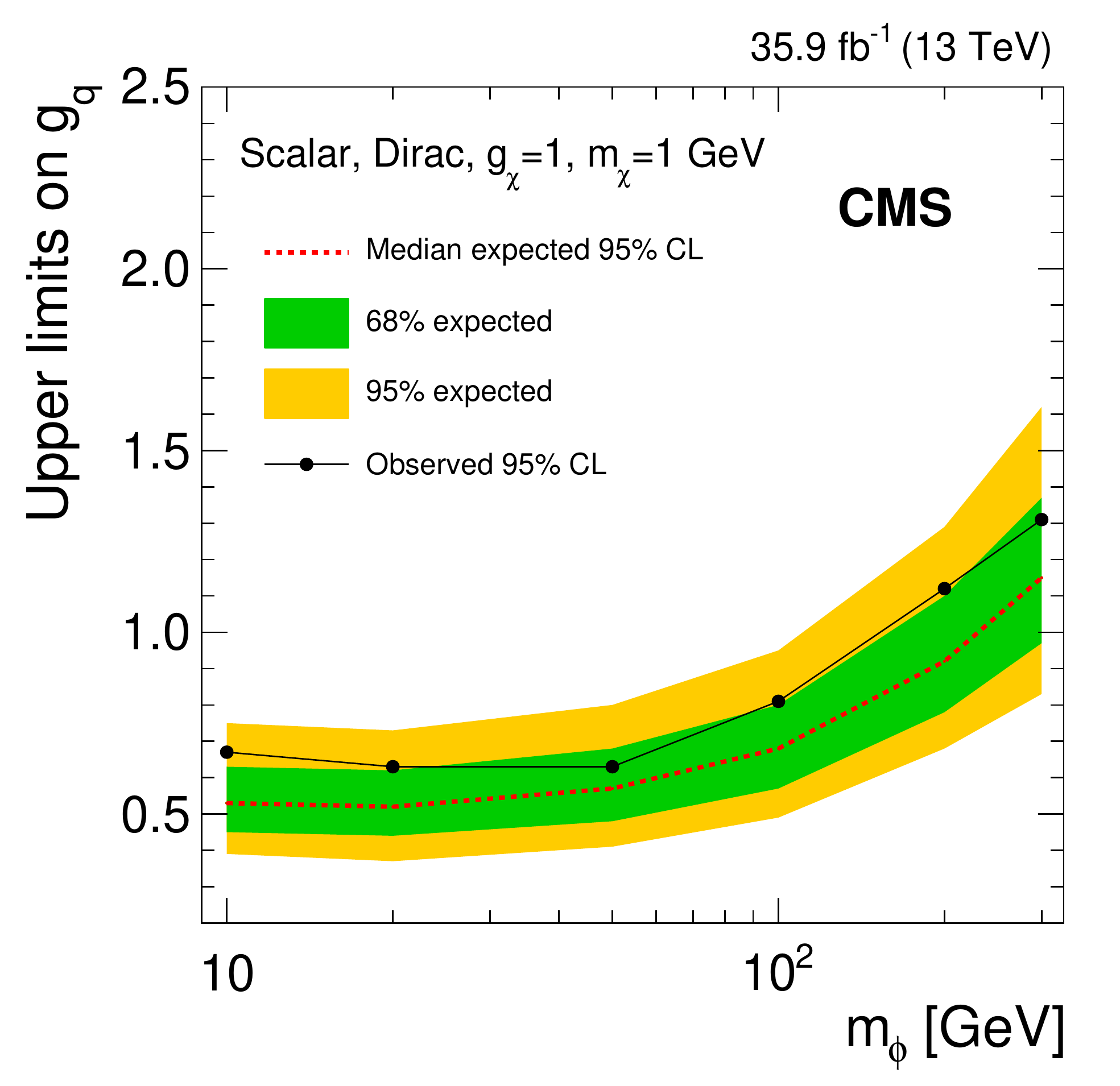}
  \includegraphics[width=0.45\textwidth]{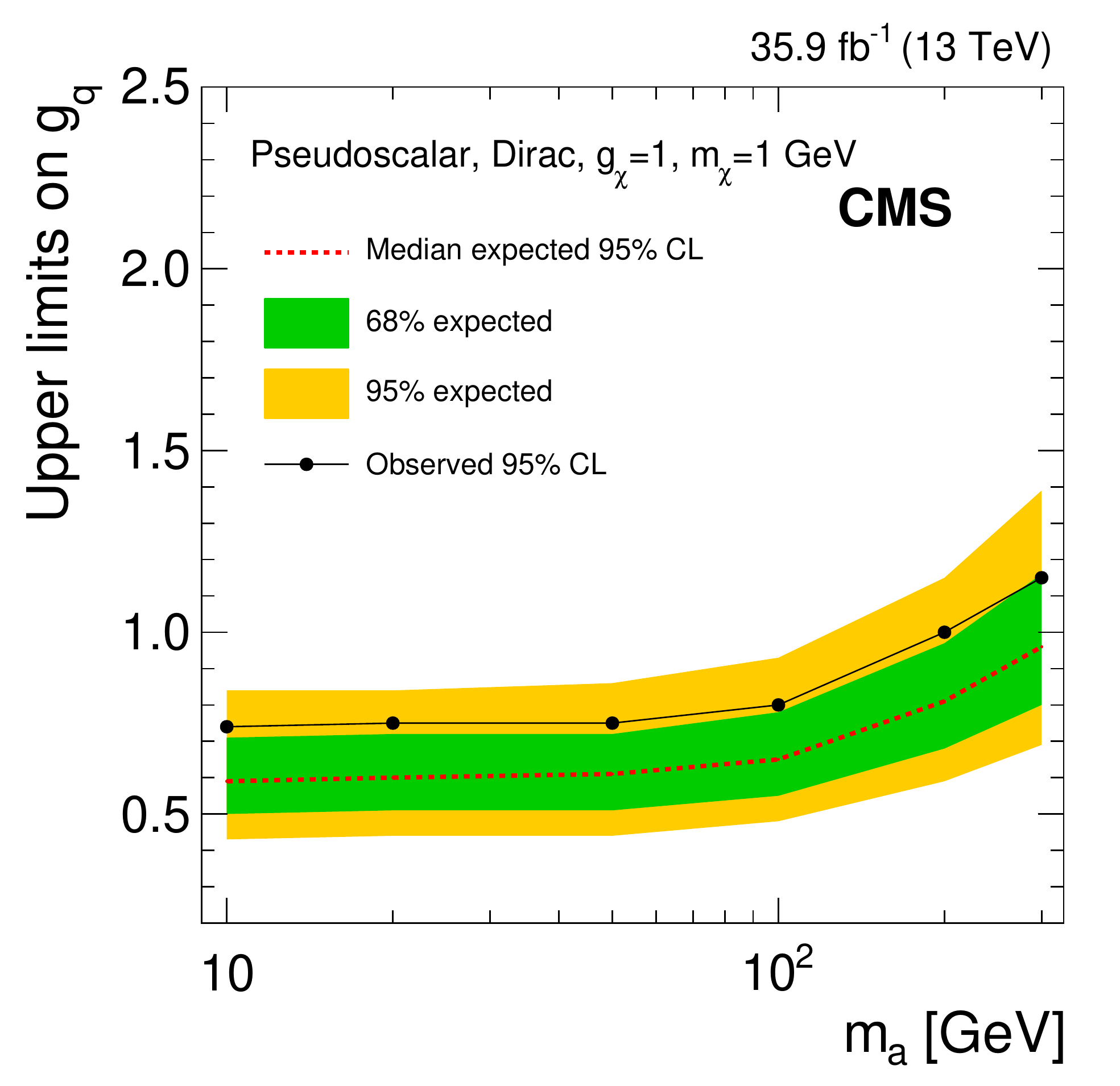}
  \caption{The 95\% observed and median expected \CL upper limits on the coupling strength of the mediator to the standard model quarks under the assumption that $\Pg_{\chi}=1$.  A dark matter particle with a mass of 1\GeV is assumed.  The green and yellow bands indicate respectively the 68\% and 95\% probability intervals around the expected limit.  The interpretations for a scalar (\cmsLeft) and a pseudoscalar (\cmsRight) mediator are shown.}
  \label{fig:couplings_mmed_combined}
\end{figure}

The limits on $\mu$ are also expressed in terms of the mediator coupling strength to quarks in Fig.~\ref{fig:couplings_mmed_combined}.  These results are obtained by fixing $m_\chi=1\GeV$ and $\Pg_\chi=1$, and then finding the value of \gq that corresponds to the upper limit on the cross section.  This procedure is valid because the kinematic properties of the signal do not vary appreciably with \gq. The width-to-mass ratio is around 4\% for the \gq and $m_{\Pa/\phi}$ values considered.

In summary, a comprehensive search for dark matter particles produced in association with a top quark pair yields no significant excess over the predicted background.  The results presented in this Letter provide 30\&--60\% better cross section limits compared to earlier searches targeting the same signature~\cite{SUS17001,ATLASCONF2017037,Aaboud2018}.  The analysis offers stronger constraints than direct and indirect experiments for dark matter masses of O($10\GeV$) and below.  Over much of the parameter space, the $\ttDM$ signature has better sensitivity for spin-0 mediators than dark matter production in association with a jet~\cite{EXO16048} -- previously considered to be the most sensitive signature.  For the pseudoscalar model, the $\ttDM$ signature provides the most stringent cross section constraints for mediator masses of around 200\GeV and below.  The observed (expected) limits exclude a pseudoscalar mediator with mass below 220 (320)\GeV under the $\gq=\Pg_{\chi}=1$ benchmark scenario.  The $\ttDM$ signature provides the best sensitivity for the scalar mediator model and is currently the only collider signature that is sufficiently sensitive to exclude regions of parameter space with these values of the couplings.  The observed exclusion of a mediator with mass below 160 \GeV (240 \GeV expected) provides the most stringent constraint to date on this model.

\begin{acknowledgments}
We congratulate our colleagues in the CERN accelerator departments for the excellent performance of the LHC and thank the technical and administrative staffs at CERN and at other CMS institutes for their contributions to the success of the CMS effort. In addition, we gratefully acknowledge the computing centers and personnel of the Worldwide LHC Computing Grid for delivering so effectively the computing infrastructure essential to our analyses. Finally, we acknowledge the enduring support for the construction and operation of the LHC and the CMS detector provided by the following funding agencies: BMWFW and FWF (Austria); FNRS and FWO (Belgium); CNPq, CAPES, FAPERJ, and FAPESP (Brazil); MES (Bulgaria); CERN; CAS, MoST, and NSFC (China); COLCIENCIAS (Colombia); MSES and CSF (Croatia); RPF (Cyprus); SENESCYT (Ecuador); MoER, ERC IUT, and ERDF (Estonia); Academy of Finland, MEC, and HIP (Finland); CEA and CNRS/IN2P3 (France); BMBF, DFG, and HGF (Germany); GSRT (Greece); OTKA and NIH (Hungary); DAE and DST (India); IPM (Iran); SFI (Ireland); INFN (Italy); MSIP and NRF (Republic of Korea); LAS (Lithuania); MOE and UM (Malaysia); BUAP, CINVESTAV, CONACYT, LNS, SEP, and UASLP-FAI (Mexico); MBIE (New Zealand); PAEC (Pakistan); MSHE and NSC (Poland); FCT (Portugal); JINR (Dubna); MON, RosAtom, RAS, RFBR and RAEP (Russia); MESTD (Serbia); SEIDI and CPAN (Spain); Swiss Funding Agencies (Switzerland); MST (Taipei); ThEPCenter, IPST, STAR, and NSTDA (Thailand); TUBITAK and TAEK (Turkey); NASU and SFFR (Ukraine); STFC (United Kingdom); DOE and NSF (USA).
\end{acknowledgments}

\ifthenelse{\boolean{cms@external}}{\clearpage}{}
\bibliography{auto_generated}

\providecommand{\href}[2]{#2}\begingroup\raggedright\begin{thebibliography}{10}%
\makeatletter
\providecommand{\hrefCMSnoop }[0]{\@secondoftwo}%
\makeatother
\providecommand{\doi}{\texttt{doi:}\begingroup \urlstyle{tt}\Url}

\bibitem{Bertone:2004pz}
\hrefCMSnoop {}{G.~Bertone, D.~Hooper, and J.~Silk, ``{Particle dark matter:
  evidence, candidates and constraints}'',} \textit{ Phys. Rept.} \textbf{ 405}
  (2005) 279,
  \href{http://dx.doi.org/10.1016/j.physrep.2004.08.031}{\doi{10.1016/j.physrep.2004.08.031}},
\href{http://www.arXiv.org/abs/hep-ph/0404175}{\texttt{arXiv:hep-ph/0404175}}.

\bibitem{Feng:2010gw}
\hrefCMSnoop {}{J.~L. Feng, ``{Dark matter candidates from particle physics and
  methods of detection}'',} \textit{ Ann. Rev. Astron. Astrophys.} \textbf{ 48}
  (2010) 495,
  \href{http://dx.doi.org/10.1146/annurev-astro-082708-101659}{\doi{10.1146/annurev-astro-082708-101659}},
\href{http://www.arXiv.org/abs/1003.0904}{\texttt{arXiv:1003.0904}}.

\bibitem{Porter:2011nv}
\hrefCMSnoop {}{T.~A. Porter, R.~P. Johnson, and P.~W. Graham, ``{Dark matter
  searches with astroparticle data}'',} \textit{ Ann. Rev. Astron. Astrophys.}
  \textbf{ 49} (2011) 155,
  \href{http://dx.doi.org/10.1146/annurev-astro-081710-102528}{\doi{10.1146/annurev-astro-081710-102528}},
\href{http://www.arXiv.org/abs/1104.2836}{\texttt{arXiv:1104.2836}}.

\bibitem{1475-7516-2018-03-026}
G.~Bertone\hrefCMSnoop {}{ {et~al.}, ``{Identifying WIMP dark matter from
  particle and astroparticle data}'',} \textit{ JCAP} \textbf{ 03} (2018) 026,
  \href{http://dx.doi.org/10.1088/1475-7516/2018/03/026}{\doi{10.1088/1475-7516/2018/03/026}},
\href{http://www.arXiv.org/abs/1712.04793}{\texttt{arXiv:1712.04793}}.

\bibitem{Haisch:2012kf}
\hrefCMSnoop {}{U.~Haisch, F.~Kahlhoefer, and J.~Unwin, ``{The impact of
  heavy-quark loops on LHC dark matter searches}'',} \textit{ JHEP} \textbf{
  07} (2013) 125,
  \href{http://dx.doi.org/10.1007/JHEP07(2013)125}{\doi{10.1007/JHEP07(2013)125}},
\href{http://www.arXiv.org/abs/1208.4605}{\texttt{arXiv:1208.4605}}.

\bibitem{Lin:2013sca}
\hrefCMSnoop {}{T.~Lin, E.~W. Kolb, and L.-T. Wang, ``{Probing dark matter
  couplings to top and bottom quarks at the LHC}'',} \textit{ Phys. Rev. D}
  \textbf{ 88} (2013) 063510,
  \href{http://dx.doi.org/10.1103/PhysRevD.88.063510}{\doi{10.1103/PhysRevD.88.063510}},
\href{http://www.arXiv.org/abs/1303.6638}{\texttt{arXiv:1303.6638}}.

\bibitem{Buckley:2014fba}
\hrefCMSnoop {}{M.~R. Buckley, D.~Feld, and D.~Gon{\c{c}}alves, ``{Scalar
  simplified models for dark matter}'',} \textit{ Phys. Rev. D} \textbf{ 91}
  (2015) 015017,
  \href{http://dx.doi.org/10.1103/PhysRevD.91.015017}{\doi{10.1103/PhysRevD.91.015017}},
\href{http://www.arXiv.org/abs/1410.6497}{\texttt{arXiv:1410.6497}}.

\bibitem{Haisch:2015ioa}
\hrefCMSnoop {}{U.~Haisch and E.~Re, ``{Simplified dark matter top-quark
  interactions at the LHC}'',} \textit{ JHEP} \textbf{ 06} (2015) 078,
  \href{http://dx.doi.org/10.1007/JHEP06(2015)078}{\doi{10.1007/JHEP06(2015)078}},
\href{http://www.arXiv.org/abs/1503.00691}{\texttt{arXiv:1503.00691}}.

\bibitem{Arina:2016cqj}
C.~Arina\hrefCMSnoop {}{ {et~al.}, ``A comprehensive approach to dark matter
  studies: exploration of simplified top-philic models'',} \textit{ JHEP}
  \textbf{ 11} (2016) 111,
  \href{http://dx.doi.org/10.1007/JHEP11(2016)111}{\doi{10.1007/JHEP11(2016)111}},
  \href{http://www.arXiv.org/abs/1605.09242}{\texttt{arXiv:1605.09242}}.

\bibitem{Aad:2015zva}
\hrefCMSnoop {}{{ATLAS Collaboration}, ``{Search for new phenomena in final
  states with an energetic jet and large missing transverse momentum in pp
  collisions at $\sqrt{s}=$ 8 TeV with the ATLAS detector}'',} \textit{ Eur.
  Phys. J. C} \textbf{ 75} (2015) 299,
  \href{http://dx.doi.org/10.1140/epjc/s10052-015-3517-3}{\doi{10.1140/epjc/s10052-015-3517-3}},
  \href{http://www.arXiv.org/abs/1502.01518}{\texttt{arXiv:1502.01518}}.
[Erratum: {\it Eur. Phys. J. C } {\bf 75} (2015) 408].

\bibitem{Khachatryan:2016mdm}
\hrefCMSnoop {}{{CMS Collaboration}, ``{Search for dark matter in proton-proton
  collisions at 8 TeV with missing transverse momentum and vector boson tagged
  jets}'',} \textit{ JHEP} \textbf{ 12} (2016) 083,
  \href{http://dx.doi.org/10.1007/JHEP12(2016)083}{\doi{10.1007/JHEP12(2016)083}},
\href{http://www.arXiv.org/abs/1607.05764}{\texttt{arXiv:1607.05764}}.

\bibitem{Aaboud:2016tnv}
\hrefCMSnoop {}{{ATLAS Collaboration}, ``{Search for new phenomena in final
  states with an energetic jet and large missing transverse momentum in pp
  collisions at $\sqrt{s}=$ 13 TeV using the ATLAS detector}'',} \textit{ Phys.
  Rev. D} \textbf{ 94} (2016) 032005,
  \href{http://dx.doi.org/10.1103/PhysRevD.94.032005}{\doi{10.1103/PhysRevD.94.032005}},
\href{http://www.arXiv.org/abs/1604.07773}{\texttt{arXiv:1604.07773}}.

\bibitem{Sirunyan2017}
\hrefCMSnoop {}{{CMS Collaboration}, ``{Search for dark matter produced with an
  energetic jet or a hadronically decaying W or Z boson at $\sqrt{s} = 13$
  TeV}'',} \textit{ JHEP} \textbf{ 07} (2017) 14,
  \href{http://dx.doi.org/10.1007/JHEP07(2017)014}{\doi{10.1007/JHEP07(2017)014}},
  \href{http://www.arXiv.org/abs/1703.01651}{\texttt{arXiv:1703.01651}}.

\bibitem{EXO16048}
\hrefCMSnoop {}{{CMS Collaboration}, ``{Search for new physics in final states
  with an energetic jet or a hadronically decaying W or Z boson and transverse
  momentum imbalance at $\sqrt{s}=13$ TeV}'',} \textit{ Phys. Rev. D} \textbf{
  97} (2018) 092005,
  \href{http://dx.doi.org/10.1103/PhysRevD.97.092005}{\doi{10.1103/PhysRevD.97.092005}},
  \href{http://www.arXiv.org/abs/1712.02345}{\texttt{arXiv:1712.02345}}.

\bibitem{Khachatryan:2015nua}
\hrefCMSnoop {}{{CMS Collaboration}, ``{Search for the production of dark
  matter in association with top-quark pairs in the single-lepton final state
  in proton-proton collisions at $\sqrt{s} = 8$ TeV}'',} \textit{ JHEP}
  \textbf{ 06} (2015) 121,
  \href{http://dx.doi.org/10.1007/JHEP06(2015)121}{\doi{10.1007/JHEP06(2015)121}},
\href{http://www.arXiv.org/abs/1504.03198}{\texttt{arXiv:1504.03198}}.

\bibitem{Aad:2014vea}
\hrefCMSnoop {}{{ATLAS Collaboration}, ``{Search for dark matter in events with
  heavy quarks and missing transverse momentum in pp collisions with the ATLAS
  detector}'',} \textit{ Eur. Phys. J. C} \textbf{ 75} (2015) 92,
  \href{http://dx.doi.org/10.1140/epjc/s10052-015-3306-z}{\doi{10.1140/epjc/s10052-015-3306-z}},
\href{http://www.arXiv.org/abs/1410.4031}{\texttt{arXiv:1410.4031}}.

\bibitem{EXO16005}
\hrefCMSnoop {}{{CMS Collaboration}, ``{Search for dark matter produced in
  association with heavy-flavor quark pairs in proton-proton collisions at
  $\sqrt{s}=13$ TeV}'',} \textit{ Eur. Phys. J. C} \textbf{ 77} (2017) 845,
  \href{http://dx.doi.org/10.1140/epjc/s10052-017-5317-4}{\doi{10.1140/epjc/s10052-017-5317-4}},
  \href{http://www.arXiv.org/abs/1706.02581}{\texttt{arXiv:1706.02581}}.

\bibitem{cmstrigger}
\hrefCMSnoop {}{{CMS Collaboration}, ``{The CMS trigger system}'',} \textit{
  JINST} \textbf{ 12} (2017) P01020,
  \href{http://dx.doi.org/10.1088/1748-0221/12/01/P01020}{\doi{10.1088/1748-0221/12/01/P01020}},
  \href{http://www.arXiv.org/abs/1609.02366}{\texttt{arXiv:1609.02366}}.

\bibitem{CMS}
\hrefCMSnoop {}{{CMS Collaboration}, ``{The CMS experiment at the CERN LHC}'',}
  \textit{ JINST} \textbf{ 3} (2008) S08004,
  \href{http://dx.doi.org/10.1088/1748-0221/3/08/S08004}{\doi{10.1088/1748-0221/3/08/S08004}}.

\bibitem{CMS-PRF-14-001}
\hrefCMSnoop {}{{CMS Collaboration}, ``{Particle-flow reconstruction and global
  event description with the CMS detector}'',} \textit{ JINST} \textbf{ 12}
  (2017) P10003,
  \href{http://dx.doi.org/10.1088/1748-0221/12/10/P10003}{\doi{10.1088/1748-0221/12/10/P10003}},
\href{http://www.arXiv.org/abs/1706.04965}{\texttt{arXiv:1706.04965}}.

\bibitem{Cacciari:2008gp}
\hrefCMSnoop {}{M.~Cacciari, G.~P. Salam, and G.~Soyez, ``{The anti-\kt jet
  clustering algorithm}'',} \textit{ JHEP} \textbf{ 04} (2008) 063,
  \href{http://dx.doi.org/10.1088/1126-6708/2008/04/063}{\doi{10.1088/1126-6708/2008/04/063}},
\href{http://www.arXiv.org/abs/0802.1189}{\texttt{arXiv:0802.1189}}.

\bibitem{Cacciari:2011ma}
\hrefCMSnoop {}{M.~Cacciari, G.~P. Salam, and G.~Soyez, ``{FastJet user
  manual}'',} \textit{ Eur. Phys. J. C} \textbf{ 72} (2012) 1896,
  \href{http://dx.doi.org/10.1140/epjc/s10052-012-1896-2}{\doi{10.1140/epjc/s10052-012-1896-2}},
\href{http://www.arXiv.org/abs/1111.6097}{\texttt{arXiv:1111.6097}}.

\bibitem{2011JInst...611002C}
\hrefCMSnoop {}{{CMS Collaboration}, ``{Determination of jet energy calibration
  and transverse momentum resolution in CMS}'',} \textit{ JINST} \textbf{ 6}
  (2011) 11002,
  \href{http://dx.doi.org/10.1088/1748-0221/6/11/P11002}{\doi{10.1088/1748-0221/6/11/P11002}},
  \href{http://www.arXiv.org/abs/1107.4277}{\texttt{arXiv:1107.4277}}.

\bibitem{Cacciari:2008gn}
\hrefCMSnoop {}{M.~Cacciari, G.~P. Salam, and G.~Soyez, ``{The catchment area
  of jets}'',} \textit{ JHEP} \textbf{ 04} (2008) 005,
  \href{http://dx.doi.org/10.1088/1126-6708/2008/04/005}{\doi{10.1088/1126-6708/2008/04/005}},
\href{http://www.arXiv.org/abs/0802.1188}{\texttt{arXiv:0802.1188}}.

\bibitem{CMS-PAS-JME-16-003}
\href {https://cds.cern.ch/record/2256875}{{CMS Collaboration}, ``{Jet
  algorithms performance in 13 TeV data}'',} CMS Physics Analysis Summary
  CMS-PAS-JME-16-003, 2017.

\bibitem{btag}
\hrefCMSnoop {}{{CMS Collaboration}, ``{Identification of heavy-flavour jets
  with the CMS detector in pp collisions at 13 TeV}'',} \textit{ JINST}
  \textbf{ 13} (2018) P05011,
  \href{http://dx.doi.org/10.1088/1748-0221/13/05/P05011}{\doi{10.1088/1748-0221/13/05/P05011}},
\href{http://www.arXiv.org/abs/1712.07158}{\texttt{arXiv:1712.07158}}.

\bibitem{Khachatryan:2015hwa}
\hrefCMSnoop {}{{CMS Collaboration}, ``{Performance of electron reconstruction
  and selection with the CMS Detector in proton-proton collisions at $\sqrt{s}=
  8$ TeV}'',} \textit{ JINST} \textbf{ 10} (2015) P06005,
  \href{http://dx.doi.org/10.1088/1748-0221/10/06/P06005}{\doi{10.1088/1748-0221/10/06/P06005}},
\href{http://www.arXiv.org/abs/1502.02701}{\texttt{arXiv:1502.02701}}.

\bibitem{Nason:2004rx}
\hrefCMSnoop {}{P.~Nason, ``{A new method for combining NLO QCD with shower
  Monte Carlo algorithms}'',} \textit{ JHEP} \textbf{ 11} (2004) 040,
  \href{http://dx.doi.org/10.1088/1126-6708/2004/11/040}{\doi{10.1088/1126-6708/2004/11/040}},
\href{http://www.arXiv.org/abs/hep-ph/0409146}{\texttt{arXiv:hep-ph/0409146}}.

\bibitem{Frixione:2007vw}
\hrefCMSnoop {}{S.~Frixione, P.~Nason, and C.~Oleari, ``{Matching NLO QCD
  computations with parton shower simulations: the POWHEG method}'',} \textit{
  JHEP} \textbf{ 11} (2007) 070,
  \href{http://dx.doi.org/10.1088/1126-6708/2007/11/070}{\doi{10.1088/1126-6708/2007/11/070}},
\href{http://www.arXiv.org/abs/0709.2092}{\texttt{arXiv:0709.2092}}.

\bibitem{Alioli:2010xd}
\hrefCMSnoop {}{S.~Alioli, P.~Nason, C.~Oleari, and E.~Re, ``{A general
  framework for implementing NLO calculations in shower Monte Carlo programs:
  the POWHEG BOX}'',} \textit{ JHEP} \textbf{ 06} (2010) 043,
  \href{http://dx.doi.org/10.1007/JHEP06(2010)043}{\doi{10.1007/JHEP06(2010)043}},
\href{http://www.arXiv.org/abs/1002.2581}{\texttt{arXiv:1002.2581}}.

\bibitem{Oleari:2010nx}
\hrefCMSnoop {}{C.~Oleari, ``{The POWHEG-BOX}'',} \textit{ Nucl. Phys. Proc.
  Suppl.} \textbf{ 205-206} (2010) 36,
  \href{http://dx.doi.org/10.1016/j.nuclphysbps.2010.08.016}{\doi{10.1016/j.nuclphysbps.2010.08.016}},
\href{http://www.arXiv.org/abs/1007.3893}{\texttt{arXiv:1007.3893}}.

\bibitem{Alwall:2014hca}
J.~Alwall\hrefCMSnoop {}{ {et~al.}, ``{The automated computation of tree-level
  and next-to-leading order differential cross sections, and their matching to
  parton shower simulations}'',} \textit{ JHEP} \textbf{ 07} (2014) 079,
  \href{http://dx.doi.org/10.1007/JHEP07(2014)079}{\doi{10.1007/JHEP07(2014)079}},
\href{http://www.arXiv.org/abs/1405.0301}{\texttt{arXiv:1405.0301}}.

\bibitem{Denner:2009gj}
\hrefCMSnoop {}{A.~Denner, S.~Dittmaier, T.~Kasprzik, and A.~Muck,
  ``{Electroweak corrections to W+jet hadroproduction including leptonic
  W-boson decays}'',} \textit{ JHEP} \textbf{ 08} (2009) 075,
  \href{http://dx.doi.org/10.1088/1126-6708/2009/08/075}{\doi{10.1088/1126-6708/2009/08/075}},
\href{http://www.arXiv.org/abs/0906.1656}{\texttt{arXiv:0906.1656}}.

\bibitem{Denner:2011vu}
\hrefCMSnoop {}{A.~Denner, S.~Dittmaier, T.~Kasprzik, and A.~Muck,
  ``{Electroweak corrections to dilepton+jet production at hadron
  colliders}'',} \textit{ JHEP} \textbf{ 06} (2011) 069,
  \href{http://dx.doi.org/10.1007/JHEP06(2011)069}{\doi{10.1007/JHEP06(2011)069}},
\href{http://www.arXiv.org/abs/1103.0914}{\texttt{arXiv:1103.0914}}.

\bibitem{Denner:2012ts}
\hrefCMSnoop {}{A.~Denner, S.~Dittmaier, T.~Kasprzik, and A.~Maeck,
  ``{Electroweak corrections to monojet production at the LHC}'',} \textit{
  Eur. Phys. J. C} \textbf{ 73} (2013) 2297,
  \href{http://dx.doi.org/10.1140/epjc/s10052-013-2297-x}{\doi{10.1140/epjc/s10052-013-2297-x}},
\href{http://www.arXiv.org/abs/1211.5078}{\texttt{arXiv:1211.5078}}.

\bibitem{Kuhn:2005gv}
\hrefCMSnoop {}{J.~H. Kuhn, A.~Kulesza, S.~Pozzorini, and M.~Schulze,
  ``{Electroweak corrections to hadronic photon production at large transverse
  momenta}'',} \textit{ JHEP} \textbf{ 03} (2006) 059,
  \href{http://dx.doi.org/10.1088/1126-6708/2006/03/059}{\doi{10.1088/1126-6708/2006/03/059}},
\href{http://www.arXiv.org/abs/hep-ph/0508253}{\texttt{arXiv:hep-ph/0508253}}.

\bibitem{Kallweit:2014xda}
S.~Kallweit\hrefCMSnoop {}{ {et~al.}, ``{NLO electroweak automation and precise
  predictions for W+multijet production at the LHC}'',} \textit{ JHEP} \textbf{
  04} (2015) 012,
  \href{http://dx.doi.org/10.1007/JHEP04(2015)012}{\doi{10.1007/JHEP04(2015)012}},
\href{http://www.arXiv.org/abs/1412.5157}{\texttt{arXiv:1412.5157}}.

\bibitem{Kallweit:2015dum}
S.~Kallweit\hrefCMSnoop {}{ {et~al.}, ``{NLO QCD+EW predictions for V+jets
  including off-shell vector-boson decays and multijet merging}'',} \textit{
  JHEP} \textbf{ 04} (2016) 021,
  \href{http://dx.doi.org/10.1007/JHEP04(2016)021}{\doi{10.1007/JHEP04(2016)021}},
\href{http://www.arXiv.org/abs/1511.08692}{\texttt{arXiv:1511.08692}}.

\bibitem{Ball:2014uwa}
\hrefCMSnoop {}{{NNPDF} Collaboration, ``{Parton distributions for the LHC Run
  II}'',} \textit{ JHEP} \textbf{ 04} (2015) 040,
  \href{http://dx.doi.org/10.1007/JHEP04(2015)040}{\doi{10.1007/JHEP04(2015)040}},
\href{http://www.arXiv.org/abs/1410.8849}{\texttt{arXiv:1410.8849}}.

\bibitem{Sjostrand:2007gs}
\hrefCMSnoop {}{T.~Sj{\"o}strand, S.~Mrenna, and P.~Z. Skands, ``{A brief
  introduction to PYTHIA 8.1}'',} \textit{ Comput. Phys. Commun.} \textbf{ 178}
  (2008) 852,
  \href{http://dx.doi.org/10.1016/j.cpc.2008.01.036}{\doi{10.1016/j.cpc.2008.01.036}},
\href{http://www.arXiv.org/abs/0710.3820}{\texttt{arXiv:0710.3820}}.

\bibitem{Khachatryan:2015pea}
\hrefCMSnoop {}{{CMS Collaboration}, ``{Event generator tunes obtained from
  underlying event and multiparton scattering measurements}'',} \textit{ Eur.
  Phys. J. C} \textbf{ 76} (2016) 155,
  \href{http://dx.doi.org/10.1140/epjc/s10052-016-3988-x}{\doi{10.1140/epjc/s10052-016-3988-x}},
  \href{http://www.arXiv.org/abs/1512.00815}{\texttt{arXiv:1512.00815}}.

\bibitem{Agostinelli:2002hh}
\hrefCMSnoop {}{{GEANT4} Collaboration, ``{\GEANTfour}---a simulation
  toolkit'',} \textit{ Nucl. Instrum. Meth. A} \textbf{ 506} (2003) 250,
\href{http://dx.doi.org/10.1016/S0168-9002(03)01368-8}{\doi{10.1016/S0168-9002(03)01368-8}}.

\bibitem{Abercrombie:2015wmb}
\hrefCMSnoop {}{D.~Abercrombie {et~al.}, ``{Dark matter benchmark models for
  early LHC Run-2 searches: report of the ATLAS/CMS Dark Matter Forum}'',}
  (2015).
\href{http://www.arXiv.org/abs/1507.00966}{\texttt{arXiv:1507.00966}}.

\bibitem{ALBERT201749}
\hrefCMSnoop {}{A.~Albert {et~al.}, ``Towards the next generation of simplified
  dark matter models'',} \textit{ Phys. Dark Univ.} \textbf{ 16} (2017) 49,
  \href{http://dx.doi.org/10.1016/j.dark.2017.02.002}{\doi{10.1016/j.dark.2017.02.002}},
  \href{http://www.arXiv.org/abs/1607.06680}{\texttt{arXiv:1607.06680}}.

\bibitem{PhysRevD.91.055009}
\hrefCMSnoop {}{P.~Harris, V.~V. Khoze, M.~Spannowsky, and C.~Williams,
  ``Constraining dark sectors at colliders: Beyond the effective theory
  approach'',} \textit{ Phys. Rev. D} \textbf{ 91} (2015) 055009,
  \href{http://dx.doi.org/10.1103/PhysRevD.91.055009}{\doi{10.1103/PhysRevD.91.055009}},
  \href{http://www.arXiv.org/abs/1411.0535}{\texttt{arXiv:1411.0535}}.

\bibitem{MT2W}
\hrefCMSnoop {}{Y.~Bai, H.-C. Cheng, J.~Gallicchio, and J.~Gu, ``{Stop the top
  background of the stop search}'',} \textit{ JHEP} \textbf{ 07} (2012) 110,
  \href{http://dx.doi.org/10.1007/JHEP07(2012)110}{\doi{10.1007/JHEP07(2012)110}},
  \href{http://www.arXiv.org/abs/1203.4813}{\texttt{arXiv:1203.4813}}.

\bibitem{Lester:1999tx}
\hrefCMSnoop {}{C.~G. Lester and D.~J. Summers, ``{Measuring masses of
  semi-invisibly decaying particles pair produced at hadron colliders}'',}
  \textit{ Phys. Lett. B} \textbf{ 463} (1999) 99,
  \href{http://dx.doi.org/10.1016/S0370-2693(99)00945-4}{\doi{10.1016/S0370-2693(99)00945-4}},
\href{http://www.arXiv.org/abs/hep-ph/9906349}{\texttt{arXiv:hep-ph/9906349}}.

\bibitem{Burns:2008va}
\hrefCMSnoop {}{M.~Burns, K.~Kong, K.~T. Matchev, and M.~Park, ``{Using
  subsystem $M_{T2}$ for complete mass determinations in decay chains with
  missing energy at hadron colliders}'',} \textit{ JHEP} \textbf{ 03} (2009)
  143,
  \href{http://dx.doi.org/10.1088/1126-6708/2009/03/143}{\doi{10.1088/1126-6708/2009/03/143}},
\href{http://www.arXiv.org/abs/0810.5576}{\texttt{arXiv:0810.5576}}.

\bibitem{Cheng:2008hk}
\hrefCMSnoop {}{H.-C. Cheng and Z.~Han, ``{Minimal kinematic constraints and
  $m_{T2}$}'',} \textit{ JHEP} \textbf{ 12} (2008) 063,
  \href{http://dx.doi.org/10.1088/1126-6708/2008/12/063}{\doi{10.1088/1126-6708/2008/12/063}},
\href{http://www.arXiv.org/abs/0810.5178}{\texttt{arXiv:0810.5178}}.

\bibitem{pdg}
\hrefCMSnoop {}{{Particle Data Group}, ``{Review of particle physics}'',}
  \textit{ Chin. Phys. C} \textbf{ 40} (2016) 100001 and 2017 update,
\href{http://dx.doi.org/10.1088/1674-1137/40/10/100001}{\doi{10.1088/1674-1137/40/10/100001}}.

\bibitem{RooStats}
L.~Moneta\hrefCMSnoop {}{ {et~al.}, ``{The RooStats Project}'',} in \textit{
  {Proceedings, 13th International Workshop on Advanced computing and analysis
  techniques in physics research (ACAT2010)}}, p.~057.
\newblock Jaipur, India, February, 2010.
\newblock \href{http://www.arXiv.org/abs/1009.1003}{\texttt{arXiv:1009.1003}}.
\newblock [PoS(ACAT2010)057].
\href{http://dx.doi.org/10.22323/1.093.0057}{\doi{10.22323/1.093.0057}}.

\bibitem{CMS-PAS-LUM-17-001}
\href {https://cds.cern.ch/record/2257069}{{CMS Collaboration}, ``{CMS}
  luminosity measurements for the 2016 data taking period'',} CMS Physics
  Analysis Summary CMS-PAS-LUM-17-001, 2017.

\bibitem{PhysRevD.95.092001}
\hrefCMSnoop {}{{CMS Collaboration}, ``{Measurement of differential cross
  sections for top quark pair production using the lepton+jets final state in
  proton-proton collisions at 13 TeV}'',} \textit{ Phys. Rev. D} \textbf{ 95}
  (2017) 092001,
  \href{http://dx.doi.org/10.1103/PhysRevD.95.092001}{\doi{10.1103/PhysRevD.95.092001}},
\href{http://www.arXiv.org/abs/1610.04191}{\texttt{arXiv:1610.04191}}.

\bibitem{JUNK1999435}
\hrefCMSnoop {}{T.~Junk, ``Confidence level computation for combining searches
  with small statistics'',} \textit{ Nucl. Instrum. Methods A} \textbf{ 434}
  (1999) 435,
  \href{http://dx.doi.org/10.1016/S0168-9002(99)00498-2}{\doi{10.1016/S0168-9002(99)00498-2}},
\href{http://www.arXiv.org/abs/hep-ex/9902006}{\texttt{arXiv:hep-ex/9902006}}.

\bibitem{cls}
\hrefCMSnoop {}{A.~L. Read, ``{Presentation of search results: the $CL_s$
  technique}'',} \textit{ J. Phys. G} \textbf{ 28} (2002) 2693,
  \href{http://dx.doi.org/10.1088/0954-3899/28/10/313}{\doi{10.1088/0954-3899/28/10/313}}.

\bibitem{Cowan2011}
\hrefCMSnoop {}{G.~Cowan, K.~Cranmer, E.~Gross, and O.~Vitells, ``Asymptotic
  formulae for likelihood-based tests of new physics'',} \textit{ Eur. Phys. J.
  C} \textbf{ 71} (2011) 1554,
  \href{http://dx.doi.org/10.1140/epjc/s10052-011-1554-0}{\doi{10.1140/epjc/s10052-011-1554-0}},
  \href{http://www.arXiv.org/abs/1007.1727}{\texttt{arXiv:1007.1727}}.
[Erratum: \DOI{10.1140/epjc/s10052-013-2501-z}].

\bibitem{SUS17001}
\hrefCMSnoop {}{{CMS Collaboration}, ``{Search for top squarks and dark matter
  particles in opposite-charge dilepton final states at $\sqrt{s}=$ 13 TeV}'',}
  \textit{ Phys. Rev. D} \textbf{ 97} (2018) 032009,
  \href{http://dx.doi.org/10.1103/PhysRevD.97.032009}{\doi{10.1103/PhysRevD.97.032009}},
\href{http://www.arXiv.org/abs/1711.00752}{\texttt{arXiv:1711.00752}}.

\bibitem{ATLASCONF2017037}
\hrefCMSnoop {}{{ATLAS Collaboration}, ``{Search for top-squark pair production
  in final states with one lepton, jets, and missing transverse momentum using
  36.1 fb$^{-1}$ of $\sqrt{s}=$ 13 TeV pp collision data with the ATLAS
  detector}'',} (2017).
  \href{http://www.arXiv.org/abs/1711.11520}{\texttt{arXiv:1711.11520}}.
Submitted to \textit{JHEP}.

\bibitem{Aaboud2018}
\hrefCMSnoop {}{{ATLAS Collaboration}, ``{Search for dark matter produced in
  association with bottom or top quarks in $\sqrt{s}=13$ TeV pp collisions with
  the ATLAS detector}'',} \textit{ Eur. Phys. J. C} \textbf{ 78} (2018) 18,
  \href{http://dx.doi.org/10.1140/epjc/s10052-017-5486-1}{\doi{10.1140/epjc/s10052-017-5486-1}},
  \href{http://www.arXiv.org/abs/1710.11412}{\texttt{arXiv:1710.11412}}.

\end{thebibliography}\endgroup
\cleardoublepage \appendix\section{The CMS Collaboration \label{app:collab}}\begin{sloppypar}\hyphenpenalty=5000\widowpenalty=500\clubpenalty=5000\vskip\cmsinstskip
\textbf{Yerevan Physics Institute, Yerevan, Armenia}\\*[0pt]
A.M.~Sirunyan, A.~Tumasyan
\vskip\cmsinstskip
\textbf{Institut f\"{u}r Hochenergiephysik, Wien, Austria}\\*[0pt]
W.~Adam, F.~Ambrogi, E.~Asilar, T.~Bergauer, J.~Brandstetter, M.~Dragicevic, J.~Er\"{o}, A.~Escalante~Del~Valle, M.~Flechl, R.~Fr\"{u}hwirth\cmsAuthorMark{1}, V.M.~Ghete, J.~Hrubec, M.~Jeitler\cmsAuthorMark{1}, N.~Krammer, I.~Kr\"{a}tschmer, D.~Liko, T.~Madlener, I.~Mikulec, N.~Rad, H.~Rohringer, J.~Schieck\cmsAuthorMark{1}, R.~Sch\"{o}fbeck, M.~Spanring, D.~Spitzbart, A.~Taurok, W.~Waltenberger, J.~Wittmann, C.-E.~Wulz\cmsAuthorMark{1}, M.~Zarucki
% ...
\vskip\cmsinstskip
% ...
1:  Also at Vienna University of Technology, Vienna, Austria\\
% ...
\end{sloppypar}
\end{document}